\documentclass[usegraphicx,usenatbib,useAMS]{mn2e}
\usepackage{times}
\usepackage{epsfig}

\begin{document}

\newcommand{\be}{\begin{equation}}
\newcommand{\ee}{\end{equation}}
\newcommand{\ba}{\begin{eqnarray}}
\newcommand{\ea}{\end{eqnarray}}
\newcommand{\mat}{{\bf}}
\newcommand{\nn}{\nonumber\\}
\newcommand{\bA}{{\bf A}}
\newcommand{\bE}{{\bf E}}
\newcommand{\bB}{{\bf B}}
\newcommand{\bC}{{\bf C}}
\newcommand{\bF}{{\bf F}}
\newcommand{\bS}{{\bf S}}
\newcommand{\bfa}{{\bf a}}
\newcommand{\bb}{{\bf b}}
\newcommand{\bg}{{\bf g}}
\newcommand{\bj}{{\bf j}}
\newcommand{\bk}{{\bf k}}
\newcommand{\bn}{{\bf n}}
\newcommand{\bp}{{\bf p}}
\newcommand{\br}{{\bf r}}
\newcommand{\bu}{{\bf u}}
\newcommand{\bv}{{\bf v}}
\newcommand{\bx}{{\bf x}}
\newcommand{\by}{{\bf y}}
\newcommand{\bbmu}{\mbox{\boldmath $\mu$}}
\newcommand{\dd}{{\partial}}
\newcommand{\ddt}{{\partial\over \partial t}}
\newcommand{\lnL}{\ln{\cal L}}
\newcommand{\bbtheta}{\mbox{\boldmath $\theta$}}
\def\gs{\mathrel{\raise1.16pt\hbox{$>$}\kern-7.0pt %  >
\lower3.06pt\hbox{{$\scriptstyle \sim$}}}}         %  ~
\def\ls{\mathrel{\raise1.16pt\hbox{$<$}\kern-7.0pt %  <
\lower3.06pt\hbox{{$\scriptstyle \sim$}}}}         %  ~

\title[Star formation and metallicity
histories of SDSS galaxies]{Star
Formation and Metallicity History of the SDSS galaxy survey: unlocking the fossil
record}
\author[Panter, Heavens \& Jimenez]
{Benjamin Panter$^1$\thanks{email: bdp@roe.ac.uk; afh@roe.ac.uk;
    raulj@physics.rutgers.edu},
Alan F. Heavens$^1$\footnotemark[1], Raul Jimenez$^2$\footnotemark[1] \\
$^1$Institute for Astronomy, Dept. of Physics \& Astronomy, Blackford Hill,
Edinburgh EH9-3HJ, UK; bdp, afh@roe.ac.uk\\
$^2$Department of Physics and Astronomy, Rutgers University, NJ-08854-8019,
USA; raulj@physics.rutgers.edu}

\maketitle

\begin{abstract}
  Using MOPED we determine non-parametrically the star-formation and
  metallicity history of over 37,000 high-quality galaxy spectra from the
  Sloan Digital Sky Survey (SDSS) early data release.  We use the entire
  spectral range, rather than concentrating on specific features, and we
  estimate the complete star formation history without prior assumptions about
  its form (by constructing so-called `population boxes').  The main results
  of this initial study are that the star formation rate in SDSS galaxies has
  been in decline for $\sim$ 6 Gyr; a metallicity distribution for
  star-forming gas which is peaked $\sim 3$ Gyr ago at about solar
  metallicity, inconsistent with closed-box models, but consistent with infall
  models. We also determine the infall rate of gas in SDSS and show
  that it has been significant for the last 3 Gyr. We investigate errors using
  a Monte-Carlo Markov Chain algorithm.  Further, we demonstrate that
  recovering star formation and metallicity histories for such a large sample
  becomes intractable without data compression methods, particularly the
  exploration of the likelihood surface.  By exploring the whole likelihood
  surface we show that age-metallicity degeneracies are not as severe as by
  using only a few spectral features.  We find that 65\% of galaxies contain a
  significant old population (with an age of at least 8 Gyr), including recent
  starburst galaxies, and that over 97\% have some stars older than 2 Gyr. It
  is the first time that a complete star formation and metallicity history,
  without restrictive assumptions about its form have been derived for such a
  large dataset of integrated stellar populations, and the first
  time that the past star formation history has been determined
  from the fossil record of the present-day spectra of galaxies.
\end{abstract}


\section{Introduction}

The measured spectrum of a galaxy contains, in principle, information
about the physical processes that led to its formation and evolution.
The amount of gas transformed into stars, the metallicity of that gas
and the dust produced in it at a given time all affect the integrated
light of a galaxy. Therefore, nearby spectra should contain a precious
fossil record about the conditions of the interstellar medium in the
past, and can be compared to methods based on measurements of recent
star formation activity, measured at different redshifts (cf
\citet{Lilly+96,Madau+96,H+98}; see also \cite{Baldry+02}).
The challenge is to recover {\em all} the information contained in
the spectrum. In principle, this is not a difficult task since the
composite spectrum of a stellar population will be just the sum of
single stellar population spectra over the history of the galaxy.
An essentially non-parametric reconstruction of the star formation
history and metallicity can be achieved by searching for the
best-fitting model. This can easily be attempted for a small
number spectra using the whole data set of measured fluxes, with
weak constraints on the star formation history, such as piecewise
continuity. Despite its simplicity, this approach has never been
used and it is most common to assume a very simple parametrisation
of the star formation (usually a declining exponential with the
amplitude and decay time as parameters) and a given metallicity
history. In addition, most of the analysis is usually done on
pre-selected features (absorption or emission lines) of the
spectrum. This limits the amount of information that can be
extracted and also introduces artificial degeneracies among the
parameters that could be lifted using all information in the
spectrum.

More sophisticated approaches to recover physical information from galaxy
spectra than simply using a pre-selected set of spectral features or
broad-band colours have been used in the literature.  Many of these are based
on principal component analysis (PCA) or wavelet decomposition (e.g.
\citet{Murtagh87,Francis92,Connolly95,Ronen99,Folkes99,MSLE02}), based on
information theory \citep{SSTL00} or solving the inverse problem
\citep*{VergelyLanconMouchine02}. However, when dealing with large datasets,
like SDSS, with potentially 10$^6$ spectra, it is impractical to use all of
the flux data for every galaxy - searching for the best fitting model for a
spectrum with 10,000 flux points takes about an hour on a high-end PC linux
work station.  Data compression of some sort is necessary, and this can be
achieved by concentrating on particular stellar features, such as the 4000\AA\
break and $H\delta$, plus broad-band colours \citep{Kauffmann+02}. The
approach of MOPED (Multiple Optimised Parameter Estimation and Data
compression; \citep*{HJL00}) is rather different.  It chooses a relatively
small number of linear combinations of the data, where the weightings are
chosen carefully and automatically to preserve as much information as possible
about the parameters one wants to know about (the star formation and
metallicity histories, and the dust content).  In this way it is possible in a
practical way to recover virtually as much information as is theoretically
possible, given the data and a theoretical model.

In this paper we use the radical data compression algorithm MOPED
to greatly reduce the time that is needed to find a best-fitting
model. This algorithm allows us to obtain an essentially
non-parametric reconstruction of the star formation and
metallicity histories of the galaxies in the early data release of
the Sloan Digital Sky Survey (SDSS).  As important as the results
obtained is the fact that the fast algorithm allows us to explore
the likelihood surface and obtain realistic errors.  We have
previously shown the potential of MOPED with a very modest sample
\citep*{RJH01} and also demonstrated its clear advantages over PCA
where a good forward model exists \citep*{HJL00}.

We have also implemented a Monte-Carlo Markov Chain (MCMC)
algorithm to explore the corresponding likelihood surface. We
describe in detail convergence criteria, step size and the length
of the chain needed to sample properly the surface. This is the
only feasible method to explore efficiently possible degeneracies
and covariances in the recovered parameters. The likelihood
surface appears to be far from gaussian, so errors computed using
local approximations (such as the Fisher matrix) may yield grossly
inaccurate error estimates. This has a profound effect on possible
correlations between parameters.

Our main findings are that the star formation rate in SDSS
galaxies has been in decline for $\sim$ 6 Gyr, with tentative
evidence for flattening after that time; a metallicity
distribution for star-forming gas which is peaked $\sim 3$ Gyr
ago, inconsistent with closed-box models, but consistent with
infall models; and a very slight correlation of dust content with
the level of current star formation.

This paper is organised as follows: in \S 2 we describe briefly
the data compression algorithm. The method to recover star
formation and metallicity histories and dust is described in \S 3
and technical details in \S 4. Main results are presented in \S 5
and \S 6 presents our conclusions. A detailed account on how to
build Markov Chains is given in an Appendix, with special
emphasis on how to choose the jump size and to decide when the
chain has converged.

\section{MOPED}

With a survey like SDSS that will contain $\sim 700,000$ spectra,
each containing a few thousand flux measurements, it becomes
almost impossible to do a brute force search on large-dimensional
likelihood surfaces. More specifically, to find the best fitting
model for each of the galaxies, each having 3850 flux
measurements, in a 25-dimensional space, would require 2 years of
CPU in a high-end Linux PC workstation. Taking into account that
one also needs to find errors in the parameters, i.e. to explore
the likelihood surface, the problem becomes effectively
intractable since, as we will show below, the number of likelihood
evaluations needed for each spectrum is quite large.  For this
paper, we actually use 300,000 steps.

Fortunately, it is not necessary to include all the flux measurements
independently in the model fitting - some of the data may tell us very
little about the parameters we are trying to estimate. This may be
because the flux measurements are not sensitive to the parameters or
they are very noisy. One obvious route to reduce the number of data
points is simply to remove them, but this is not optimal in general
and some information will be lost.  A more fruitful route is to
construct linear combinations of the data with weightings chosen
carefully to avoid losing information about the star formation and
metallicity history.  In \citet*{HJL00} such a method was developed
and it was later termed MOPED.  Remarkably, MOPED reduces the size of
the dataset to a compressed dataset comprising one datum per
parameter, without losing information provided certain conditions are
met.  A priori, it is by no means obvious that this can be done.

The advantage of such a method to tackle the above problem is
obvious, since now the time taken to calculate the likelihood is
reduced by the ratio of the number of original data points to the
number of parameters.  In cases of correlated data, such as in the
microwave background power spectrum, the acceleration is even
larger - the cube of this ratio \citep{GuptaHeavens02}. Clearly
this efficient method of determining the star formation history is
invaluable when dealing with large spectral datasets, such as the
SDSS, but it also opens up the possibility of describing the star
formation and metallicity history in a relatively free-form way,
not restricted to simple parametrisations.  In this paper, we
describe the star formation and metallicity history by values in
12 bins of look-back time, spaced logarithmically, and we also
estimate one dust parameter. Determining 25 parameters per galaxy
would be impractical without MOPED.

The method is as follows.  Given a set of data {\bf x} (in our
case the spectrum of a galaxy) which includes a signal part
${\bbmu}$ and noise ${\bf n}$, i.e. $\bx = \bbmu + \bn$, the idea
then is to find weighting vectors ${\bf b}_m$ such that $y_m
\equiv {\bf b}_m^{t} {\bf x}$ contain as much information as
possible about the parameters (star formation rates, metallicity
etc.).  These {\it numbers} $y_m$ are then used as the data set in
a likelihood analysis.  In MOPED, there is one vector associated
with each parameter.

In \citet*{HJL00} an optimal and lossless method was found to calculate
${\bf b}_m$ for multiple parameters (as is the case with galaxy
spectra).  The definition of lossless here is that the Fisher matrix
at the maximum likelihood point (see \citet*{TTH97}) is the same
whether we use the full dataset or the compressed version.  The Fisher
matrix gives a good estimate of the errors on the parameters, provided
the likelihood surface is well described by a multivariate Gaussian
near the peak.  The method is strictly lossless in this sense provided
that the noise is independent of the parameters, and provided our
initial guess of the parameters is correct.  This is not exactly true
for galaxy spectra, owing to the presence of a shot noise component
from the source photons, and because our initial guess is inevitably
wrong.  However, the increase in parameter errors is very small in
these cases (see \citet*{HJL00}) - MOPED recovers the correct solutions
extremely accurately even when the conditions for losslessness are not
satisfied.  The weights required are
\begin{equation}
\bb_1 = {\bC^{-1} \bbmu_{,1}\over \sqrt{\bbmu_{,1}^t
\bC^{-1}\bbmu_{,1}}}
\label{Evector1}
\end{equation}
and
\begin{equation}
\bb_m = {\bC^{-1}\bbmu_{,m} - \sum_{q=1}^{m-1}(\bbmu_{,m}^t
\bb_q)\bb_q \over
\sqrt{\bbmu_{,m}^t \bC^{-1} \bbmu_{,m} - \sum_{q=1}^{m-1}
(\bbmu_{,m}^t \bb_q)^2}} \qquad (m>1).
\label{bbm}
\end{equation}
where a comma denotes the partial derivative with respect to the
parameter $m$ and $C$ is the covariance matrix with components
$C_{ij}=\langle n_in_j\rangle$. $m$ runs from 1 to the number of
parameters $M$, and $i$ and $j$ from 1 to the size of the dataset
(the number of flux measurements in a spectrum).  To compute the
weight vectors requires an initial guess of the parameters.  We
term this the fiducial model.

The dataset $\{y_m\}$ is orthonormal: i.e. the $y_m$ are
uncorrelated, and of unit variance. The new likelihood is easy to
compute (the $y_m$ have means $\langle y_m \rangle = \bb_m^t
\bbmu$), namely: \be \ln{\cal L}(\theta_\alpha) = {\rm constant} -
\sum_{m=1}^{M} {(y_m-\langle y_m\rangle)^2\over 2}. \ee Further
details are given in \citet*{HJL00}.

It is important to note that if the covariance matrix is known for a
large dataset (e.g. a large galaxy redshift survey) or it does not
change significantly from spectrum to spectrum, then the $\langle y_m
\rangle$ need be computed only {\em once} for the whole dataset, thus
with massive speed up factors in computing the likelihood as will be
shown in \S 3 and \S 4.  Note that the $y_m$ are only orthonormal
if the fiducial model coincides with the correct one.  In practice one
finds that the recovered parameters are almost completely independent
of the choice of fiducial model, but one can iterate if desired to
improve the solution.

\section{PARAMETRISATION OF STAR FORMATION, METALLICITY AND DUST}

\begin{figure*}
\includegraphics[width=17cm,height=14cm]{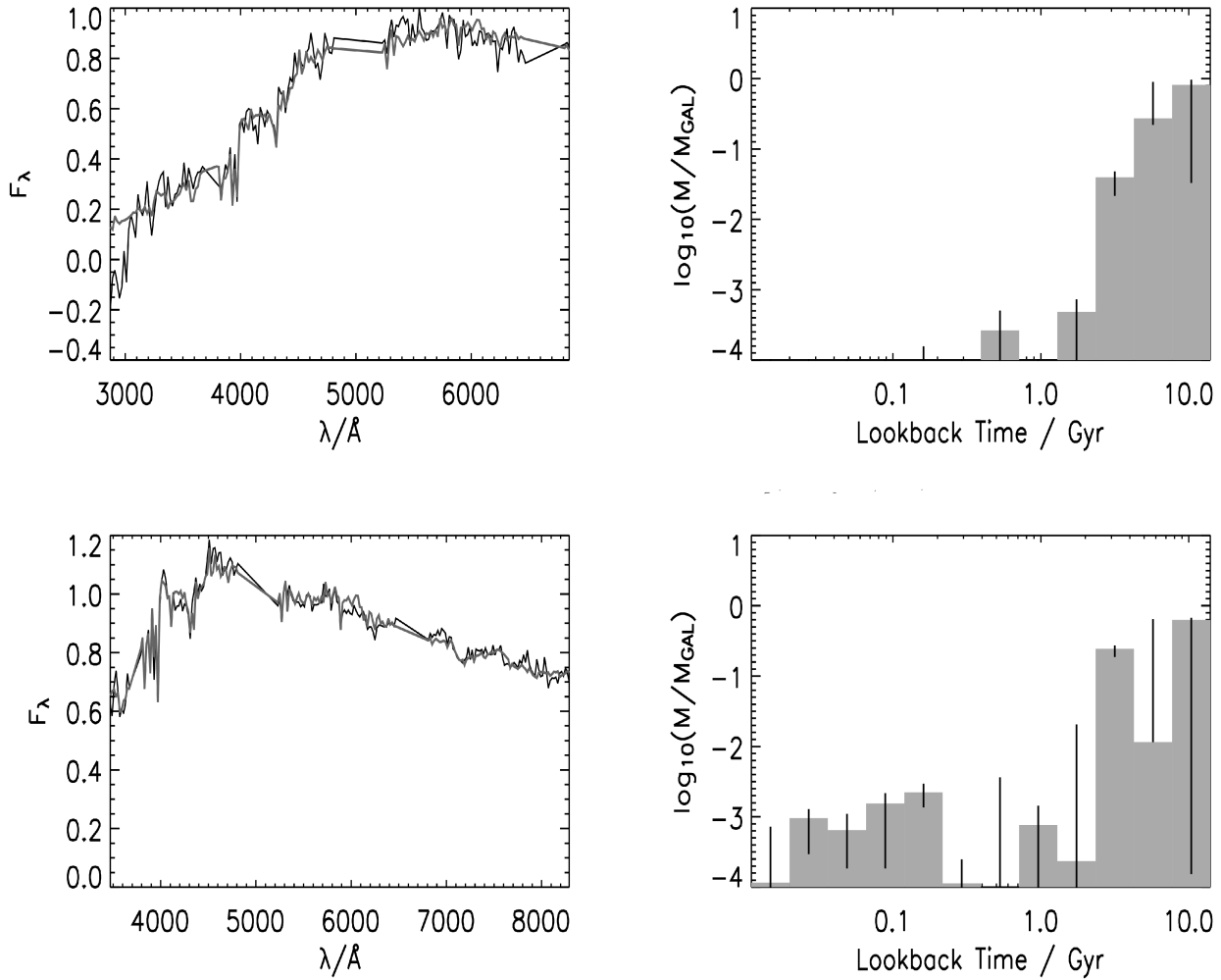}
\caption{Two examples of spectra with the best-fitting model found by
MOPED (left panels).  The right-hand panels show the recovered star
formation history, along with marginal errors determined by the
MCMC method (see Appendix for details). Three areas have been removed
to avoid contamination of the fitted spectrum by emission lines, at
wavelengths around 3800, 5000 and 6600 angstroms.  These regions are
ignored in the fitting.  The SDSS identifications of the galaxies are
shown above the figures.}
\label{fig:popboxes}
\end{figure*}
Star formation in galaxies takes place in giant molecular clouds that
are relatively short-lived (about $10^7$ yr or less) during the whole
life of the galaxy.  It is therefore clear that star formation can be
described in a rather model independent way by dividing time into
widths of $10^7$ yr, each of which has a given metallicity. Then for a
galaxy today whose star formation extends over most of the age of the
universe, the problem amounts to determining $\approx 10^3$
parameters. This is firstly not tractable with current computing power
and furthermore, as we will demonstrate below, the observed spectra of
current galaxies may not have sensitivity to all episodes of star
formation. Thus we adopt a different strategy which consists on a
coarser grid than above. The bins in (lookback) time are chosen to
have equal width in logarithmic space (this is discussed in detail in
\citet*{RJH01}). With logarithmic lookback times, the error in the
final spectrum caused by the uncertainty of the exact time at which
star formation in the bin occured is roughly independent of time. The
large bin widths at early times simply reflect the fact that there is
little sensitivity in the final spectrum to the exact time that star
formation takes place. 

In the section below we will show that star formation histories and
metallicities can be recovered with sensible errors if the grid
contains 12 bins.  The age bins start at a lookback time of 0.01,
increasing in equal logarithmic steps with a spacing of 0.258.  To
each of these bins we assign two numbers, the fraction of the total
stellar mass created in that time bin and the metallicity of the gas
which formed that mass. Dust is modeled in a simple way by use of the
\citet{C97} colour excess parameter, E(B-V), sufficient to describe
the major effect of dust absorption on the integrated light of
galaxies. The Calzetti model depends only on one parameter: the amount
of dust in the galaxy. It is obvious that more sophisticated models
are needed to describe the effect of dust in galaxies and this will be
explored in a future paper. The model effectivly tilts the spectrum by
supressing the blue end. The set of simple stellar population models
used in this paper is the one by \citet{JPMH98,JDMPP03}, and we refer
the reader to those references for a thorough discussion of the
validity of the models (see also \citet*{JFK98,KFJ02}).  Line emission
from the galaxies has been removed, as the stellar spectral synthesis
models do not include emission lines from gaseous regions, and the
relationship between the emission line strengths and star formation
history is less certain than the modeling of stellar features.  A
technicality is that the rest-frame wavelength coverage of the
galaxies obviously depends on the redshift of the galaxy, so we make a
one-off computation of a different set of MOPED vectors $\bb_m$ for
each of 49 values of redshift between $z=0$ and $0.34$.  Extending the
redshift range requires further MOPED vectors to be computed, and
there are very few galaxies beyond $z=0.34$.

For reference, the fiducial model has star formation which increases
linearly for increasing bin age, as well as the metallicity. The dust
parameter is 0.02. The noise is assumed to be uncorrelated, and each
wavelength bin considered is assumed to have the same error.  This is
not a bad approximation for the Sloan galaxies, after emission lines
have been removed.  The level of the noise simply scales the MOPED
vectors, and we use the appropriate average error for each galaxy.  We
emphasize though that the precise choice of fiducial model and
covariance matrix is not crucial and does not bias the parameter
estimation, nor does a poor choice lead to significantly worse errors.

\section{SDSS, and computational details}

The SDSS includes $\sim 2$\AA\ spectroscopy and $u, g, r, i$ and $z$
photometry, and will eventually contain $\sim$700,000 objects.  We
have analysed 37,752 galaxies from the Early Data Release
\citep{EDR02}, after removing objects not classified as galaxies or
outside our redshift range.  Details of the survey can be found in
\citet{Gunn+98,York+00,Strauss+02}.  The spectra were binned to 20\AA\
resolution, to match the models.  Examples are shown in
Fig.\ref{fig:popboxes}.

It is not a trivial task to determine the best-fitting parameters,
as the problem shows certain near-degeneracies (the well-known
age-metallicity degeneracy being one), and the parameter space to
be searched is large (25-dimensional).  In addition, no guaranteed
method exists to find a global maximum of the likelihood surface.
We use a two-stage process for this task.  First, a conjugate
gradient method is used, with 50 random starting points, to reduce
the chance of finding only local maxima which are not the global
maximum.  Second, the MCMC method is used to find the shape of the
likelihood surface and determine errors; details are given in the
Appendix. Finding formal errors is not a trivial task, for two
reasons. Firstly, the reduced $\chi^2$ values of the best fits are
formally too large (typically around six), which reflects the fact
that spectrophotometric modelling in not perfect. In this case,
the standard method of error determination (eg. Press et al. 1992)
fails. The second problem is that, for noisy spectra (see  
fig. 1) degeneracies do remain, and it is difficult to
quote errors for a multi-peaked likelihood surface. We assign
errors by allowing $\Delta\chi^2$(full)=number of parameters. This
procedure generally characterises the width of the highest peak
in the likelihood surface.

The MCMC stage in some cases will find a better
maximum likelihood solution, and this is then used for the
parameter estimates.  Typically, the conjugate gradient stage
takes 40 seconds per galaxy, and the MCMC stage, with 300,000
evaluations, about 3 minutes per galaxy, on a 1.6 GHz Athlon PC
workstation.  Fig.\ref{fig:popboxes} shows the recovered star
formation fractions and corresponding metallicities in the 12
lookback time periods for four galaxies.  Note that the scale is
logarithmic, and note that the mass of stars created in each time
bin is normalised to the total mass of the {\em best-fitting
model}.  Thus the error bars can extend above a fraction of unity.
We see that the star formation fractions are determined reasonably
accurately where there is a significant contribution, but
inevitably poorly determined otherwise.  12 bins is about the
maximum number justified by the data; covariances between the
estimates of the star formation in adjacent bins are beginning to
appear. This is supported by information-theoretic studies of SDSS
galaxy spectra which indicate $\sim 20$ independent components (J.
Riden, private communication). This can be appreciated in the
third panel from the top. There is a strong covariance between the
estimates in the last two bins. However, this is for a galaxy with
a very noisy spectrum. If the spectrum has got a higher signal to
noise, as in the top panel, the covariance decreases.

Although Fig.~\ref{fig:popboxes} only shows two examples, they are
representative of general trends in our analysis.  The top galaxy
shows a relatively old stellar population, with little star formation
in the last Gyr.  The lower galaxy shows a galaxy with evidence of a
higher level of more recent star formation.  Some remaining degeneracy
is apparent in the results, especially in the oldest two bins, where
some trade-off between the two is allowed.  For the lower galaxy,
although the formal best fit puts almost all of the old stars in the
oldest bin, the error bars indicate that there are solutions which are
almost as good which transfer much of that star formation into the
6Gyr old bin.  Population boxes for the whole sample studied here will
be freely available through {\tt http://www.roe.ac.uk} in due course.

\section{Results}

\subsection{Global star formation history of galaxies}

By recovering the star formation and metallicity history of so
many galaxies, we are able to derive the global histories.  The
assumption which we make here is that galaxies which are not
selected for the SDSS have, on average, the same historical
properties as the average of those which are. It is clear, though,
that at high-redshift we are preferentially selecting bright
galaxies, compared to low redshift (SDSS is a magnitude-limited
survey).

\begin{figure}
\centerline{\psfig{file=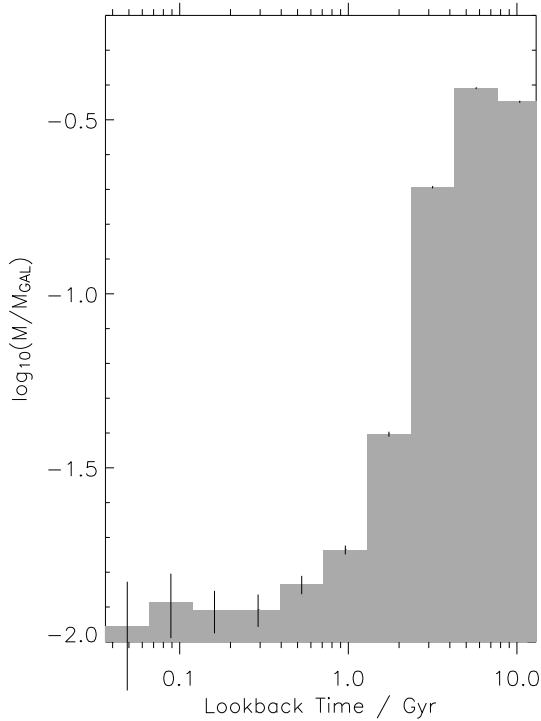,width=8.0cm,angle=0,clip=}}
\caption{The average mass fraction of gas converted into stars as a
function of time and lookback time.  This figure is weighted by
number.}
\label{fig:SFH}
\end{figure}

\begin{figure}
\centerline{\psfig{file=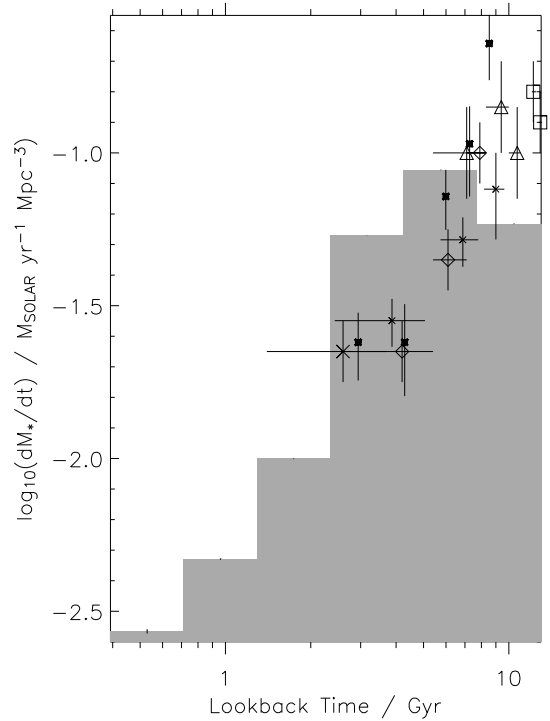,width=8.0cm,angle=0,clip=}}
\caption{The rate of conversion of gas to stars, plotted against
lookback time; results are weighted by the stellar mass.  The symbols
indicate previous measurements: diamonds are from \citet{Lilly+96},
triangles from \citet{Connolly+97}, squares from \citet{SAGDP99} and
the cross from \citet{TresseMaddox98}. All points have been dust
corrected using the \citet{SAGDP99} correction factor.}
\label{fig:dMdt}
\end{figure}

The first quantity of interest is the mass fraction of gas converted into
stars as a function of redshift. This is shown in Fig. \ref{fig:SFH}.  Star
formation fractions have been determined for each galaxy as a function of
rest-frame galaxy lookback time.  The measurements are then assigned to
lookback time relative to the present day by rebinning.  We see that about 1/3
of the star formation in the SDSS galaxies occurs in the last 4 Gyr, and
another 1/3 was formed more than 8 Gyr ago. The quality of the SDSS spectra
does not seem to allow a finer time binning for old ages than the one
presented here (see covariances between bins in Fig.~\ref{fig:popboxes}), so
it is difficult to estimate how many were formed at a really high redshift.
Using currently-favoured cosmological parameters (flat universe,
$\Omega_m=0.3$, $H_0=72$ km s$^{-1}$ Mpc$^{-1}$), a lookback time of 8 Gyr
corresponds only to $z \sim 1$. It is clear that moderately higher signal to
noise spectra would provide with the possibility of a finer grid of the oldest
bin and thus a more accurate determination of the rate at which gas is
converted into stars at high redshift. However, the recent bins (ages smaller
than 4 Gyr, or equivalently $z < 0.4$) are well resolved. It is interesting
that 30\% is also the fraction of stars presently in spheroids, which contain
the oldest stellar populations, thus we could infer that most stars in
spheroids were formed more than $8$ Gyr ago.

A quantity of great interest for the past few years has been the
volume-average star formation rate in the universe as a function of
redshift (e.g. \citet{Lilly+96,Madau+96,H+98,SAGDP99}). This is
derived by determining the {\em current} star formation rate from
star-forming galaxies at different redshifts.  Fig. \ref{fig:dMdt}
shows our findings from the SDSS fossil record evidence.  As is
apparent from the figure, we find a similar sharp decline in star
formation rate with time, but we are able to extend the star formation
histories to much more recent times, and we see that the trend
continues at least to lookback times of 0.5 Gyr, representing a drop
by a factor of 30 or more.  The formal errors suggest a real increase
in star formation at early times, but we caution against
overinterpretation here in view of degeneracies in the last two bins
which are apparent in fig. \ref{fig:popboxes}. Further, we have
assumed that galaxies below the magnitude limit of the SDSS have the
same star formation rate. Although, in the bins of fig. \ref{fig:dMdt}
we only plot the statistical errorsfrom the MOPED fitting, we estimate
that other systematic errors (small number of galaxies and the exact
number of galaxies per $Mpc^3$) contribute about 30\% to the error in
the height of each bin. The overplotted points are some recent
measurements of the current star formation rate of galaxies from
different surveys using the compilation in \citet{SAGDP99}.
Specifically, diamonds are from \citet{Lilly+96}, triangles from
\citet{Connolly+97} and squares from \citet{SAGDP99}. All points have
been dust corrected using the \citet{SAGDP99} correction factor. In
addition we also plotted (cross) the measurement at $z \sim 0.2$ by
\citet{TresseMaddox98}.

The overall shape of the SFR as recovered from the SDSS is in
reasonably good agreement with the instantaneous SFR estimates.  We do
find that the oldest bin is slightly below the penultimate one
indicating a star formation rate of about 35\% lower at early times
than at the peak.  However, we have found that these last two bins
often show degeneracies in individual spectra, so the turndown at
early times may not be significant.  We also find that the decline in star
formation continues to small lookback times.

\subsection{Metallicity evolution with redshift}

\begin{figure}
\centerline{\psfig{file=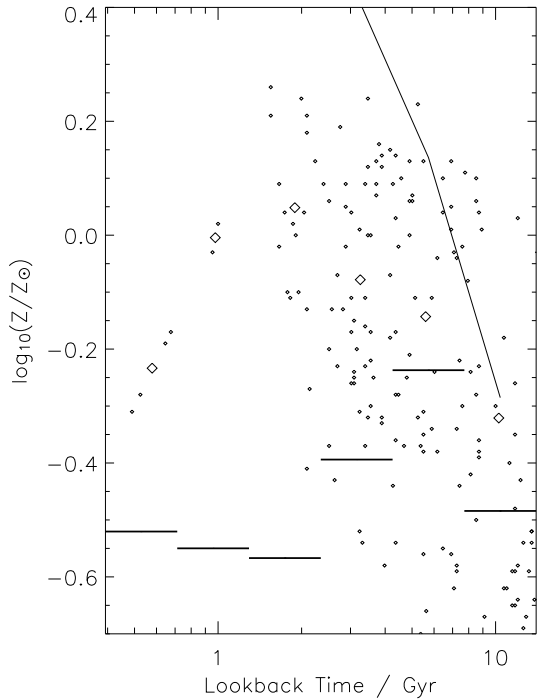,width=8.0cm,angle=0,clip=}}
\caption{The average value of the metallicity for the 37,752 galaxies studied
  in this paper (horizontal bars).  Also shown are stars from the Milky Way,
  and their averages (small and large diamonds) and the predictions of a
  closed-box model (solid line).}
\label{fig:metals}
\end{figure}

\begin{figure}
\centerline{\psfig{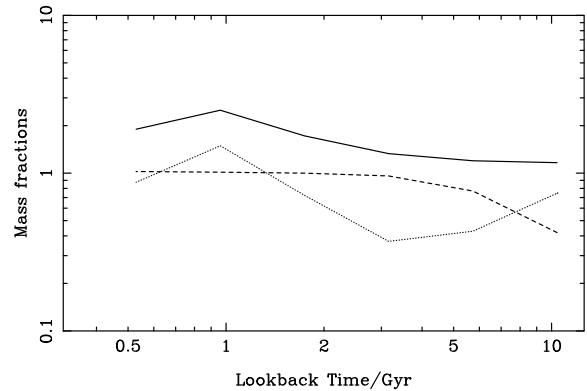}}
\caption{The evolution of the average total mass (solid), stellar mass
(dashed) and gas mass (dotted), normalised to the final stellar mass,
for an infall model constrained to give the right star formation and
metallicity histories.}
\label{fig:acc}
\end{figure}

\begin{figure}
\centerline{\psfig{file=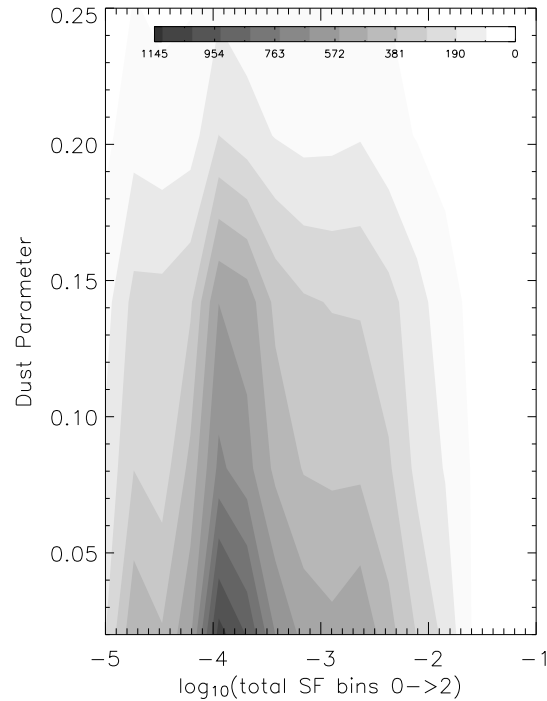,width=8.0cm,angle=0,clip=}}
\caption{The joint distribution of dust content and recent (last 40
Myr) star formation rate in the SDSS sample. The number of galaxies in
each contour level is shown in the upper bar.} \label{fig:dust}
\end{figure}

We also compute the average metallicity of the gas which turns
into stars at each epoch.  This is shown in Fig.\ref{fig:metals},
and shows some striking features.  At high redshift the average
metallicity increases with time, but then shows a systematic
decrease, providing strong support for infall of relatively
unprocessed gas into galaxies at lookback times between 4 and 0.1
Gyr. The average metallicity peaks at values of about solar. While
stars formed more than 8 Gyr ago and less than 2 Gyr have
metallicities slightly lower, about half solar. The rise in
metallicity from early times to age $\sim 3$ Gyr can be accounted
for by closed-box models, but not the strong decline in
metallicity at lookback times of less than 3 Gyr: the solid line
shows the predictions of the closed-box model (e.g.
\citet{BinneyMerrifield98})
\begin{equation}
Z(t)=-p\, ln \left [ \frac{M_g(t)}{M_g(0)} \right ]
\end{equation}
where p is the yield, $M_g(t)$ the mass in gas at time $t$ and
$M_g(0)$ the initial gas mass.  Note that for ages below $3$ Gyr,
the closed-box model is inconsistent with the data. The line has
$p=0.02$ and $M_g(0)=1$ (in units of the final stellar mass), but
it is obvious that no closed-box model can account for the decline
of metallicity.

The diamond symbols correspond to ages and metallicities from the
\citet{Edvardsson+93} sample of stars in the Milky Way with
accurate ages and metallicities.  The large diamonds average these
data. Although the SDSS galaxies include spheroids and not only
disk systems, it is gratifying that the overall shape and range of
$Z(t)$ is similar to that for Milky Way stars, although the Milky
Way is apparently offset to later times when compared with the
SDSS population.

Fig. \ref{fig:acc} shows the results from an accretion-box or
infall model (e.g.  \citet{BinneyMerrifield98}), where we allow
for fresh infall of gas constrained to reproduce the star
formation and metallicity history. The total mass (stars plus gas)
$M_t$ then obeys the following equation
\begin{equation}
\delta M_t = {p\over Z}\delta M_s - (M_t - M_s) {\delta Z\over Z}.
\end{equation}
The figure assumes $p=0.02$ and $M_g(0)=1$ (normalised to the
final stellar mass), but the general features of the curve are
robust, with fresh infall clearly necessary to reduce the
metallicity. The different lines correspond to the evolution of
the average total mass (solid), stellar mass (dashed) and gas mass
(dotted), normalised to the final stellar mass, for an infall
model constrained to give the right star formation and metallicity
histories. Note the significant amount of gas supply in the last 3
Gyr.   With these yield and initial gas mass, outflow is formally
required in the last Gyr.

\subsection{Dust content}

In Fig. \ref{fig:dust} we show the joint distribution of dust
content and star formation fraction in the last 40 Myr.  There is
no strong evidence of a trend of higher dust content with higher
recent star formation, in contrast to the findings of
\citet{Kauffmann+02}.

\section{Discussion and Conclusions}

We have determined the past star formation history of the
Universe from the present-day fossil record of the spectra of
more-or-less present-day galaxies, and made comparisons with
previous alternative methods based on computing the essentially
instantaneous star formation rate from samples over a range of
redshifts.  We have presented a MOPED analysis of 37,752 galaxy
spectra from the SDSS Early Data Release.  MOPED allows very
rapid determination of the star formation and metallicity history
of each galaxy, through carefully-designed and optimised data
compression.  We have therefore been able to dispense with the
usual oversimplifying assumptions which are usually employed to
make this sort of problem tractable, such as
exponentially-decaying star formation rates, bursts and so on. We
recover the star formation fraction and the metallicity in 12
equal size bins of log(lookback time), plus one parameter
describing the dust content, using the \citet{C97} model.

Previous studies of the star formation history of galaxies have
generally used measurements of the contemporary star formation
rates,  and the history has been constructed by making
observations at a range of redshifts.  These studies have
indicated a decline in the star formation rate to the present-day.
This behaviour should be apparent in the fossil record of the
spectrum of galaxies at low redshift.  In this paper one of our
main conclusions is that this evidence is present in the Sloan
Digital Sky Survey Early Data Release sample: the average star
formation rate in the SDSS sample has been in decline for the last
$\sim 6$ Gyr or so, at which point there is some evidence for a
peak in the star formation rate.  We find that the subsequent decline
has contintued to very recent lookback times ($<0.5$ Gyr).

On average, each galaxy produced about 30\% of the stars more than
$\sim 8$ Gyr ago. This corresponds to all stars in spheroids being
formed at $z > 1$. This age distribution is consistent with the
age distribution of stars in the Milky Way.   65\% of galaxies have
some star formation older than 8 Gyr, and 97\% have some star
formation older than 2 Gyr.

In addition, we find that the average metallicity of the gas rises
with time to a peak $\sim 3$ Gyr ago, and has been in significant
decline since then. This is clearly in contradiction with
closed-box chemical evolution models, but can be accounted for
with infall models (e.g. \citet{BinneyMerrifield98}). The maximum
average metallicity is about solar, for stars formed 2-8 Gyr ago.
Stars formed recently, $ < 1$ Gyr ago, in SDSS galaxies have
metallicities as low as $\sim 0.5 Z_{\odot}$ on average, similar
to that of stars in SDSS galaxies formed more than 8 Gyr ago. We
also find a very weak correlation between the dust content of a
galaxy and its recent star formation rate.

We can compare our analysis and results with the studies of
\citet{Kauffmann+02}. That paper used the 4000\AA\ break, the
$H\delta$ feature, and four broad-band measurements.  Thus each
galaxy spectrum was compressed into six numbers, chosen on the
basis of knowledge of the evolution of stellar spectral features
and model evolution of broad-band colours.  MOPED's approach is
rather different; firstly it uses the entire spectrum, and uses
model predictions to find which wavelengths are most sensitive to
the parameters of interest.  It then does an automatic data
compression step, reducing the spectrum to (in this paper) 25
linear combinations which retain as much information as
possible.  In this way, we are able to get results which are in
principle as accurate as the modelling allows. Furthermore,  the
speed advantage offered by the data compression step allows us to
be ambitious in what we extract: we are able to estimate star
formation fractions in twelve time bins, plus twelve associated
metallicities of the star-forming gas, and a dust parameter.
Without the radical data compression, the searching of this
25-dimensional space would be prohibitively slow, and we would be
restricted to simple parametrisations of the star formation
history, as has been done historically.

Population boxes for the whole sample studied here will be released
through {\tt http://www.roe.ac.uk} in due course.

\section*{acknowledgments}
RJ is supported by NSF grant AST-0206031.  RJ is grateful to Licia Verde and
Hiranya Peiris for many insightful discussions about Markov chains.  We thank
Chris Williams and Jamie Riden for useful discussions.

Funding for the creation and distribution of the SDSS Archive has been
provided by the Alfred P. Sloan Foundation, the Participating Institutions,
the National Aeronautics and Space Administration, the National Science
Foundation, the U.S.  Department of Energy, the Japanese Monbukagakusho, and
the Max Planck Society. The SDSS Web site is {\tt http://www.sdss.org/}.

The SDSS is managed by the Astrophysical Research Consortium (ARC) for the
Participating Institutions. The Participating Institutions are The University
of Chicago, Fermilab, the Institute for Advanced Study, the Japan
Participation Group, The Johns Hopkins University, Los Alamos National
Laboratory, the Max-Planck-Institute for Astronomy (MPIA), the
Max-Planck-Institute for Astrophysics (MPA), New Mexico State University,
University of Pittsburgh, Princeton University, the United States Naval
Observatory and the University of Washington.

%\bibliographystyle{mn2e.bst}
%\bibliography{../STY/raul,../STY/general}

\begin{thebibliography}{}

\bibitem[\protect\citeauthoryear{{Baldry et al.}}{{Baldry et
  al.}}{2002}]{Baldry+02}
{Baldry et al.} I.~K.,  2002, ApJ, 569, 582

\bibitem[\protect\citeauthoryear{{Binney} \& {Merrifield}}{{Binney} \&
  {Merrifield}}{1998}]{BinneyMerrifield98}
{Binney} J.,  {Merrifield} M.,  1998, {Galactic astronomy}.
Princeton University Press, 1998.

\bibitem[\protect\citeauthoryear{{Blain}, {Smail}, {Ivison}, {Kneib} \&
  {Frayer}}{{Blain} et~al.}{2002}]{Blain+02}
{Blain} A.~W.,  {Smail} I.,  {Ivison} R.,  {Kneib} J.-P.,    {Frayer} D.~T.,
  2002, Phys. Rept., 369, 111

\bibitem[\protect\citeauthoryear{{Calzetti}}{{Calzetti}}{1997}]{C97}
{Calzetti} D.,  1997, AJ, 113, 162

\bibitem[\protect\citeauthoryear{Connolly, Szalay, Bershady, Kinney \&
  Calzetti}{Connolly et~al.}{1995}]{Connolly95}
Connolly A.,  Szalay A.,  Bershady M.,  Kinney A.,    Calzetti D.,  1995, AJ,
  110, 1071

\bibitem[\protect\citeauthoryear{{Connolly}, {Szalay}, {Dickinson}, {Subbarao}
  \& {Brunner}}{{Connolly} et~al.}{1997}]{Connolly+97}
{Connolly} A.~J.,  {Szalay} A.~S.,  {Dickinson} M.,  {Subbarao} M.~U.,
  {Brunner} R.~J.,  1997, ApJ, 486, L11

\bibitem[\protect\citeauthoryear{{Edvardsson}, {Andersen}, {Gustafsson},
  {Lambert}, {Nissen} \& {Tomkin}}{{Edvardsson} et~al.}{1993}]{Edvardsson+93}
{Edvardsson} B.,  {Andersen} J.,  {Gustafsson} B.,  {Lambert} D.~L.,  {Nissen}
  P.~E.,    {Tomkin} J.,  1993, A\&A, 275, 101

\bibitem[\protect\citeauthoryear{{Folkes et al.}}{{Folkes et
  al.}}{1999}]{Folkes99}
{Folkes et al.} 1999, MNRAS, 308, 459

\bibitem[\protect\citeauthoryear{Francis, Hewett, Foltz \& Chaffee}{Francis
  et~al.}{1992}]{Francis92}
Francis P.,  Hewett P.,  Foltz C.,    Chaffee F.,  1992, ApJ, 398, 476

\bibitem[\protect\citeauthoryear{Gilks, Richardson \& Spiegelhalter}{Gilks
  et~al.}{1996}]{GRS96}
Gilks W.,  Richardson S.,    Spiegelhalter D.,  1996, Markov {C}hain {M}onte
  {C}arlo in {P}ractice.
Chapman and Hall

\bibitem[\protect\citeauthoryear{{Gunn et al.}}{{Gunn et al.}}{1998}]{Gunn+98}
{Gunn et al.} J.~E.,  1998, AJ, 116, 3040

\bibitem[\protect\citeauthoryear{{Gupta} \& {Heavens}}{{Gupta} \&
  {Heavens}}{2002}]{GuptaHeavens02}
{Gupta} S.,  {Heavens} A.~F.,  2002, MNRAS, 334, 167

\bibitem[\protect\citeauthoryear{Hastings}{Hastings}{1970}]{Has70}
Hastings W.,  1970, Biometrika, 57, 97

\bibitem[\protect\citeauthoryear{{Heavens}, {Jimenez} \& {Lahav}}{{Heavens}
  et~al.}{2000}]{HJL00}
{Heavens} A.,  {Jimenez} R.,    {Lahav} O.,  2000, MNRAS, 317, 965

\bibitem[\protect\citeauthoryear{{Hughes et al.}}{{Hughes et al.}}{1998}]{H+98}
{Hughes et al.} D.~H.,  1998, Nature, 394, 241

\bibitem[\protect\citeauthoryear{{Jimenez}, {Dunlop}, {MacDonald}, {Padoan} \&
  {Peacock}}{{Jimenez} et~al.}{2003}]{JDMPP03}
{Jimenez} R.,  {Dunlop} J.,  {MacDonald} J.,  {Padoan} P.,    {Peacock} J.,
  2003, in preparation

\bibitem[\protect\citeauthoryear{{Jimenez}, {Flynn} \& {Kotoneva}}{{Jimenez}
  et~al.}{1998}]{JFK98}
{Jimenez} R.,  {Flynn} C.,    {Kotoneva} E.,  1998, MNRAS, 299, 515

\bibitem[\protect\citeauthoryear{{Jimenez}, {Padoan}, {Matteucci} \&
  {Heavens}}{{Jimenez} et~al.}{1998}]{JPMH98}
{Jimenez} R.,  {Padoan} P.,  {Matteucci} F.,    {Heavens} A.~F.,  1998, MNRAS,
  299, 123

\bibitem[\protect\citeauthoryear{{Kauffmann et al.}}{{Kauffmann et
  al.}}{2002}]{Kauffmann+02}
{Kauffmann et al.} G.,  2002, astro-ph/0205070

\bibitem[\protect\citeauthoryear{{Kotoneva}, {Flynn} \& {Jimenez}}{{Kotoneva}
  et~al.}{2002}]{KFJ02}
{Kotoneva} E.,  {Flynn} C.,    {Jimenez} R.,  2002, MNRAS, 335, 1147

\bibitem[\protect\citeauthoryear{{Lilly}, {Le Fevre}, {Hammer} \&
  {Crampton}}{{Lilly} et~al.}{1996}]{Lilly+96}
{Lilly} S.~J.,  {Le Fevre} O.,  {Hammer} F.,    {Crampton} D.,  1996, ApJ, 460,
  L1

\bibitem[\protect\citeauthoryear{{Madau}, {Ferguson}, {Dickinson},
  {Giavalisco}, {Steidel} \& {Fruchter}}{{Madau} et~al.}{1996}]{Madau+96}
{Madau} P.,  {Ferguson} H.~C.,  {Dickinson} M.~E.,  {Giavalisco} M.,  {Steidel}
  C.~C.,    {Fruchter} A.,  1996, MNRAS, 283, 1388

\bibitem[\protect\citeauthoryear{{Madgwick}, {Sommerville}, {Lahav} \&
  {Ellis}}{{Madgwick} et~al.}{2002}]{MSLE02}
{Madgwick} D.,  {Sommerville} R.,  {Lahav} O.,    {Ellis} R.,  2002,
  astro-ph/0210471

\bibitem[\protect\citeauthoryear{Metropolis, Rosenbluth, Rosenbluth, Teller \&
  Teller}{Metropolis et~al.}{1953}]{MRR+53}
Metropolis N.,  Rosenbluth A.,  Rosenbluth M.,  Teller M.,    Teller E.,  1953,
  J. Chem. Phys., 21, 1087

\bibitem[\protect\citeauthoryear{Murtagh \& Heck}{Murtagh \&
  Heck}{1987}]{Murtagh87}
Murtagh F.,  Heck A.,  1987, Multivariate Data Analysis.
Reidel, Dordrecht

\bibitem[\protect\citeauthoryear{{Press}, {Teukolsky}, {Vetterling} \&
  {Flannery}}{{Press} et~al.}{1992}]{Press+92}
{Press} W.~H.,  {Teukolsky} S.~A.,  {Vetterling} W.~T.,    {Flannery} B.~P.,
  1992, {Numerical recipes in FORTRAN. The art of scientific computing}.
Cambridge: University Press, 1992, 2nd ed.

\bibitem[\protect\citeauthoryear{{Reichardt}, {Jimenez} \&
  {Heavens}}{{Reichardt} et~al.}{2001}]{RJH01}
{Reichardt} C.,  {Jimenez} R.,    {Heavens} A.~F.,  2001, MNRAS, 327, 849

\bibitem[\protect\citeauthoryear{Ronen, Aragon-Salamanca \& Lahav}{Ronen
  et~al.}{1999}]{Ronen99}
Ronen R.~T.,  Aragon-Salamanca A.,    Lahav O.,  1999, MNRAS, 303, 284

\bibitem[\protect\citeauthoryear{{Slonim}, {Somerville}, {Tishby} \&
  {Lahav}}{{Slonim} et~al.}{2000}]{SSTL00}
{Slonim} N.,  {Somerville} R.,  {Tishby} N.,    {Lahav} O.,  2000, astro-ph,
  0005306

\bibitem[\protect\citeauthoryear{{Steidel}, {Adelberger}, {Giavalisco},
  {Dickinson} \& {Pettini}}{{Steidel} et~al.}{1999}]{SAGDP99}
{Steidel} C.~C.,  {Adelberger} K.~L.,  {Giavalisco} M.,  {Dickinson} M.,
  {Pettini} M.,  1999, ApJ, 519, 1

\bibitem[\protect\citeauthoryear{{Stoughton et al.}}{{Stoughton et
  al.}}{2002}]{EDR02}
{Stoughton et al.} C.,  2002, AJ, 123, 485

\bibitem[\protect\citeauthoryear{{Strauss et al.}}{{Strauss et
  al.}}{2002}]{Strauss+02}
{Strauss et al.} M.~A.,  2002, AJ, 124, 1810

\bibitem[\protect\citeauthoryear{Tegmark, Taylor \& Heavens}{Tegmark
  et~al.}{1997}]{TTH97}
Tegmark M.,  Taylor A.,    Heavens A.,  1997, ApJ, 480, 22

\bibitem[\protect\citeauthoryear{{Tresse} \& {Maddox}}{{Tresse} \&
  {Maddox}}{1998}]{TresseMaddox98}
{Tresse} L.,  {Maddox} S.~J.,  1998, ApJ, 495, 691

\bibitem[\protect\citeauthoryear{Vergely, Lancon \& Mouchine}{Vergely
  et~al.}{2002}]{VergelyLanconMouchine02}
Vergely J.-L.,  Lancon A.,    Mouchine M.,  2002, astro-ph/0209018

\bibitem[\protect\citeauthoryear{{York et al.}}{{York et al.}}{2000}]{York+00}
{York et al.} D.,  2000, AJ, 120, 1579

\end{thebibliography}

\section*{Appendix: Error estimation through Monte Carlo Markov-Chain
  algorithms}

Given that the number of parameters will be usually large (typically
more than 10 and in this case 25), computation of a grid to explore
the likelihood surface is impractical -- simply using 10 grid points
per dimension would require $10^{25}$ evaluations. An alternative
approach is to use the Fisher matrix around the maximum of the
likelihood to compute errors. Although this is a fast and efficient
method, it assumes that the likelihood surface is a multivariate
gaussian which may not be the case in general.

\begin{figure*}
\includegraphics[width=17cm,height=16cm]{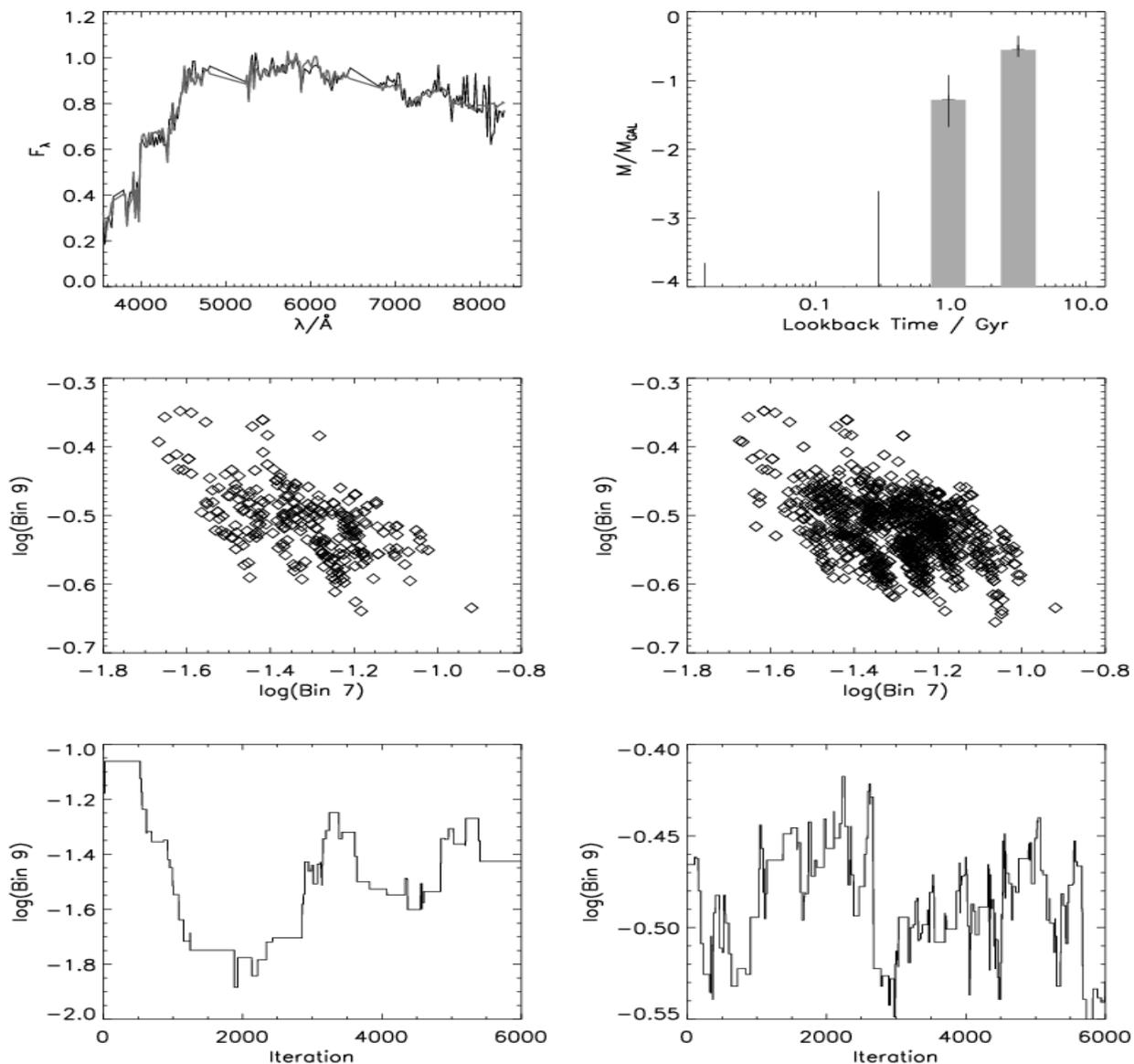}
\caption{Some details of the application of MOPED to an individual
SDSS galaxy.  The top left panel shows the galaxy spectrum (black)
and the best fit Jimenez model (grey).  The corresponding star
formation fractions are shown in the top right panel.  The middle
panels show the MCMC points with highest likelihood values, in
projection onto parameter plane 7 and 9, which are the bins
showing significant star formation, for chains of 30,000 (left)
and 300,000 (right). The lower panels show the effects of
different step sizes on the estimation of parameter 9.  The
convergence of the parameter estimates shown in the lower part of
the figure is discussed in the appendix.} \label{fig:Individual}
\end{figure*}

In the general case, an efficient method to sample the likelihood
surface is through the Markov Chain Monte Carlo (MCMC) algorithm
\citep{MRR+53,Has70}\footnote{An excellent account of Markov
chains techniques can be found in \citet{GRS96}}. In essence the
Markov chain algorithm is very simple: a chain of likelihood
values in the parameter space is created in the following way.  At
each point, a random step is made in parameter space, and a random
number between 0 and 1 is drawn.  This number is essentially
compared with the ratio of the likelihood values between the
current step and the previous one, although there are some slight
modifications, especially near the boundaries of the parameter
space. If the value of the likelihood ratio is bigger than 1 or
the random number, then the current step is accepted and added to
the chain. If, however, it is smaller than the random number,
then the point is rejected and not added to the chain.
Asymptotically, the distribution of points in the chain samples
the likelihood surface in an unbiased way.

In this paper, we use a uniform prior within certain bounds to
produce the step. If knowledge about the shape of the surface is
known a priori, then the random stepping can be done more
efficiently by using this a priori information.  The Fisher matrix
can sometimes be useful for this, but if the topology of the
likelihood hyper-surface is not known, it may be inaccurate.
Marginal errors are then trivial to compute by looking at the
distribution of all points for a single parameter. The method is
very fast and efficient but the challenge is to adjust the step
size of the jump so the likelihood surface around the maximum is
explored with the minimum number of steps.

There are some rules to decide the size of the jump a priori (see
\citet{GRS96}), we find that the most efficient time step can be
found by exploration of a few thousand chains for a few galaxy
spectra.

\begin{figure*}
\includegraphics[width=18cm,height=18cm]{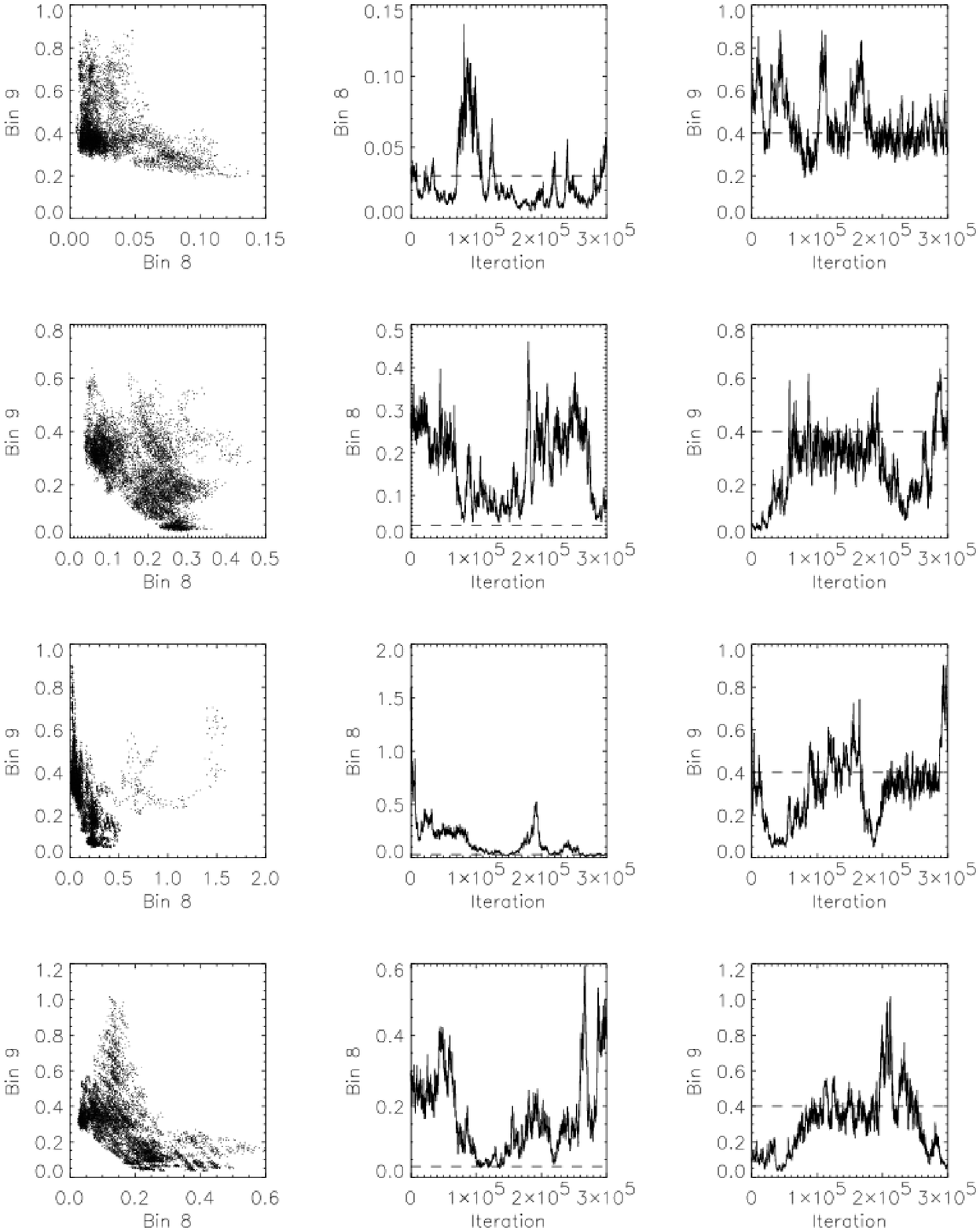}
\caption{Convergence test for MCMC. Four chains of length 300,000
steps have been started at different points in parameter space
for the galaxy shown in Fig.~\ref{fig:Individual}. The far left
panel shows the values of the star formation in bins 8 and 9 for
each chain.  In order to show how the chain samples the
likelihood surface, the middle left panel shows those points in
the chain within a reduced $\Delta\chi^2=1$ of the best-fitting
solution. The right-hand panels show the values of the star
formation parameters explored by the chains.  Note that the
excursion to the right in the third row is in fact the starting
point of the chain.  The long-dashed horizontal lines show the
best-fitting solution.} \label{fig:4chains}
\end{figure*}

As an example of Markov-chain convergence examine
Fig.~\ref{fig:Individual}. The top-left panel shows a typical
spectrum of a galaxy with an old stellar population. The top-right
panel displays the recovered star formation history with errors
computed using the MCMC. The middle-left panel shows the
distribution of points in the plane of parameter values recovered
from the MCMC for only 30,000 points in the chain, showing only
those within reduced $\Delta\chi^2=1$ of the minimum $\chi^2$
point. The middle-right panel shows the same but this time for a
chain with 300,000 points.  The chain with a small number of
points shows a rugged pattern, with a few unexplored areas, but
the chain with 300,000 points has covered the region of the
likelihood space that is most favoured quite well. More
specifically, it is clear that the chain with a small number of
points has not come back to the starting point of the chain a few
times.  This feature is required to establish convergence, and
the left hand chain is said to be not well mixed, although in
this case it provides a good estimate of the errors. On the
contrary, the right hand chain  with a small step has oscillated
a few times around the starting point.

In the above chains the time step was chosen by trial and error.
In principle, the step size can be chosen optimally a priori. For
one parameter, for example, the step size should be such that the
rejection rate of points is about 60\%, leading to a
non-negligible chance of the chain exploring regions in the
likelihood surface that are more than 3$\sigma$ away from the
best solution. Obviously, for more parameters the rejection rate
will decrease significantly, since there are many more ways the
jump can explore an unlikely region of the parameter space. High
acceptance rates are indicative of too small a jump step (see
below).

The two bottom panels of figure~\ref{fig:Individual} show the
values of one parameter (star formation for the bin with most
star formation) for part of two chains.  The left-bottom panel
shows a chain with a step that is too big. Note how the chain
remains at same value of the parameter for many steps. The path
of the parameter value for the change shows a clear ``staircase''
pattern, and the chain is inefficient.  On the other hand, the
right-bottom panel shows a chain with a much better step size. In
this case the chain does not dwell for long on a single value of
the parameter.

Our experiments show that significant improvements in chain
convergence can be obtained by using a nonuniform jump size.  The star
formation history is divided into two sections, the first covering
mass fractions from $10^{-7}$ (essentially zero) to $10^{-4}$, the
second from $10^{-4}$ to 10.  In the first region we have 100
logarithmically-spaced points, in the second a thousand.  The maximum
jump we allow is 20 steps. Similarly, there are 64 values of
metallicity in a grid, log spaced between $0.01 < Z/Z_{\odot} < 5$ and
the optimal jump size is 5 grid points. Dust is computed on a linear
grid, with 64 elements, between 0.02 and 1.28 and optimal jump size of
3 grid elements.  The chain is well mixed for 300,000 steps and the
acceptance rate is typically about 2\% in 25 dimensions.

A more robust way to estimate the convergence of the chain is the
following: start 4 or more chains from widely-separated points in
the parameter space and check when the variance for all
parameters within the chain and between chains are
indistinguishable; at this point the chains have converged. The
point is well illustrated in Fig.~\ref{fig:4chains}. The left
four panels show the distribution of points in a projection of
the likelihood hyper-surface for two adjacent bins for a random
galaxy with only significant star formation in bins 8 and 9.  The
second column of panels shows only those points for which reduced
$\Delta \chi^2 < 1$ from the best value in the likelihood, while
the 1 column of panels shows all 300,000 points in the chains. In
the right panels we show the values of bins 8 and 9 for the four
chains and 300000 jumps. The dashed line is the best-fitting
value for the parameter. A few features become apparent by visual
inspection. When the chains start from values different from the
best fit, it takes 50-100,000 steps for them to converge. After
this, the chain remains on the good valley solution for some
time, before undergoing a random excursion away from the best
solution, before returning at some point.  This returning
behaviour is required for convergence.

If we use the above convergence criterion, we see that chains
converge at different points in path. For example, the chain on the
top panel for bin 9 only converges after 200,000 points. While the
second chain from the top, does so after only 50,000 steps for bin
9. This illustrates how important it is to run chains for a long
enough time and estimate convergence.

An alternative approach is to run only one chain from one point in the
parameter space\footnote{Using the conjugate gradient method it is
easy to choose this point as the one closest to the best solution} and
let it run for long enough as to explore most of the likelihood space.
A good test to check convergence then is to monitor the likelihood
values as a function of the step.  The chain should return to close to
the maximum likelihood solution.  In the present paper we follow this
criterion and in all cases we find chains have converged after 300,000
steps, as illustrated in the example of Fig.~\ref{fig:4chains}.

It is worth emphasizing that the chain needs to be able to explore
the likelihood hyper-surface well in the vicinity of the peak in
order to be sure errors are derived properly. Also, there is some
danger in using local approximations to the likelihood surface
(the Fisher matrix) to compute the length of each jump. Imagine, a
flat valley in the likelihood surface with a narrow region within
where the likelihood is better. A first jump may bring you into
the valley, then the Fisher matrix will indicate that the
hyper-surface is locally very flat which will systematically
provide a very large jump and therefore the better value will be
systematically missed.

\subsection{Covariances between bins}

The left panels in Fig.~\ref{fig:4chains} illustrate how
covariances appear between adjacent bins.  The far left panel
shows the values of the whole chain, whereas the middle-left
panel shows only those points within reduced $\Delta\chi^2=1$ of
the best-fitting solution.  Note how the paths for bin values
follow a common pattern: star formation in one bin is traded by
star formation on the adjacent bin (with different metallicity) -
the sum of bins 8 and 9 is well constrained in this example.  We
also see from the right-hand panels that the chains spend longer
periods close to the best-fitting parameter where the parameter
has more star formation (parameter 9) than less (parameter 8).

\end{document}